\documentclass[10pt]{iopart}

\usepackage{graphicx}
\usepackage{siunitx}
\usepackage{enumerate}

\begin{document}
\maketitle
\ioptwocol
\vspace*{-1.65cm}\title[]{\normalsize Mechanisms of rapid evolution}
\newcommand*{\TitleFont}{%
    \usefont{\encodingdefault}{\rmdefault}{b}{n}%
    \fontsize{10}{10}%
    \selectfont}
\author{\TitleFont Hong-Yan Shih$^{1,2}$ and Nigel Goldenfeld$^1$}
\address{\small $^1$ Department of Physics and Carl R. Woese Institute for Genomic Biology, University of Illinois at Urbana-Champaign, Champaign, Illinois 61801, USA}
\address{\small $^2$ Institute of Physics, Academia Sinica, Taipei 11529, Taiwan}
\vspace{10pt}

%





%

\noindent\textit{Status}

Evolution is increasingly being recognized as a central component of
two of the most urgent societal problems
\cite{cavicchioli2019scientists}.  While the physics of global climate
change is relatively well-understood, the response of the biosphere is
harder to predict, especially the extent to which microbial evolution
can influence the feedback between soil and marine systems to the
carbon and other biogeochemical cycles
\cite{hutchins2017microorganisms,monroe2018ecoevolutionary,abs2019microbial}.
Even the sign of this feedback effect is hard to assess.  The emerging
world-wide health crisis due to the unexpectedly rapid evolution and
proliferation of antibiotic resistant strains of pathogenic bacteria is
our second example, one that underscores how imperfectly we understand
the mechanisms of evolution \cite{von2016dissemination}. In fact, it is
even the case that climate change can accentuate the problem of
antibiotic resistance \cite{cavicchioli2019scientists}.

Evolution is often thought of as the product of two independent classes
of process: (1) The generation of  mutations; (2) The dynamics within a
fixed environment that selects and ultimately conveys genetic variation
to fixation or dominance in a population. This narrative assumes a
separation of timescales between (1) and (2) but neglects the fact that
many ecosystems, especially those with microbes, show rapid genetic
adaptivity through strong selective stress arising either from
environmental conditions or antagonistic predation
\cite{rainey1998,buckling2007}. The resulting phenotypic diversity
\cite{rainey1998} contains individuals with new traits that in some
cases have been documented to further induce new links or forms of
interaction with others \cite{vetsigian2011}. For a constant driving
force, either a chemical potential difference across the ecosystem, or
a constant flux of energy, the resulting long time dynamics is an
ecosystem in a non-equilibrium steady state \cite{vetsigian2011},
characterized by constant change and the generation of new niches
\cite{laland2016introduction}, as opposed to one that is in a static
equilibrium steady state, characterized by a fixed community structure.
In other systems, such as methanogenic bioreactors
\cite{fernandez1999stable} and the global ocean microbiome
\cite{louca2016decoupling}, it is known that there is a non-equilibrium
steady state, characterized by a constant production rate, but constant
taxonomic turnover, suggesting the emergence of a collective metabolism
for the community. Whether or not there is a phase transition between
these two classes of stationary states as a function of driving force
is an interesting but unresolved fundamental question. When the
interactions between ecology and evolution are strong enough, such that
the evolutionary timescale is comparable to the ecological timescale,
qualitatively new phenomena arise: rapid and successive emergence of
evolved traits interfere with the ecosystem, resulting in significant
changes in population dynamics and spatiotemporal patterns.

The purpose of this roadmap article is to draw attention to two recent
highly simplified examples of these phenomena, which are sometimes
called {\it rapid evolution}.  The first focuses primarily on
population dynamics: anomalies in population cycles can reflect the
influence of strong selection and the interplay with mutations
(standing variation or de novo).  The second focuses primarily on the
way in which ecological structure can potentially be influenced by what
is arguably the most powerful source of genetic novelty: horizontal
gene transfer (HGT). Our understanding of the role of HGT in shaping
ecosystems, and vice versa, is in its infancy, but we now have the
tools to begin to not only understand these phenomena but to ask the
pertinent question of how one manages such dynamic ecosystems.  It is
well-documented that HGT as well as population flow is central to the
antibiotic resistance crisis \cite{von2016dissemination} and one would
expect that it plays a role in the biological response to climate
change. To achieve a full understanding of rapid evolution in all its
manifestations will require a concerted experimental and theoretical
effort.

\medskip
\noindent\textit{Anomalous population dynamics in rapid
evolution}

The first example focuses on anomalous population dynamics due to rapid
evolution.  The anomalous dynamics is characterized by abnormal phase
relationships and periodicity in population cycles. Certain
predator-prey ecosystems systems, such as rotifer-algae
\cite{yoshida2003} and phage-bacteria ecosystems
\cite{bohannan1997effect}, exhibit a $\pi$ phase difference between the
time series of predator and prey populations, together with a longer
period for their population cycle, as opposed to the typical
predator-prey phase difference of $\pi/2$. This abnormal phase
difference is associated with the emergence of a mutant prey which has
a defense against the predator but at some metabolic cost (so-called
{\it evolutionary cycle}). What is more bizarre is that in some
systems, following a mutation, the phase difference disappears as the prey
population becomes almost constant in time while the predator
population still oscillates but with a longer period than before the
mutation arose (the so-called {\it cryptic cycle}). Whether or not the
evolutionary cycle or cryptic cycle occurs depends on the metabolic
cost of defense \cite{shih2014path}.

Due to the necessarily small numbers of mutants at least during the
initial stages of these processes, one must properly take into account
the discreteness of populations and their spatial
extent\cite{durrett1994importance}.  The mathematical tools to do this
properly use stochastic individual-level models to describe the
interactions between members of the ecosystem, and statistical
mechanics techniques to deduce the resulting dynamics at the population
level \cite{mckane2005predator,butler2009robust}.

By including the trade-off between selection on reproduction and the
metabolic cost of defense against predation, a minimal stochastic
individual-level model \cite{shih2014path} reproduces the rapid
evolution that was found in chemostat experiments of rotifer-algae
\cite{yoshida2003} and phage-bacteria ecosystems
\cite{bohannan1997effect}. Under strong predation selection, a defended
prey can arise from mutation, causing the population dynamics to
transition from the normal cycle with a $\pi/2$ phase shift to the
evolutionary cycle with a $\pi$ phase shift between the predator and
the total prey.  The additional $\pi/2$ phase delay comes from the fact
that the wild-type prey, which is mostly consumed by the predator, can
only grow back after the defended sub-population starts to decrease due
to the depletion of food. When the metabolic cost of the defended prey
is low enough, the regrowth of the wild-type prey is delayed more.  If
the delay is so great that the wild-type prey can only resume growth
after the defended prey population has decreased sufficiently, the phase
delay of the wild-type prey behind the defensive prey can become $\pi$
so that the total prey population looks almost constant with time,
leading to a cryptic cycle.

The individual-level model shows that the anomalous dynamics can arise
from demographic noise in rapid evolution, without special assumptions
or fine tuning. Deterministic models are problematic, as they can not
even capture regular population cycles in a qualitative way without
introducing phenomena extra to the Lotka-Volterra description, such as
functional response \cite{mckane2005predator}.  So far, we have focused
primarily on well-mixed systems.  However, in practice, one may be
interested in invasion fronts, regime shifts or range expansion.  In
these cases, the need for correct treatment of demographic
stochasticity is even greater, because of the presence of fronts where
the populations are necessarily small. The study of the potentially
interesting spatio-temporal patterns \cite{butler2009robust} forming in
rapid evolution is a rich topic for future work.\newline

\medskip
\noindent\textit{Collective rapid evolution}

Our second example is from marine microbial ecology, and involves a
case where spatio-temporal dynamics emerges from the eco-evolutionary
feedback at various scales. In such cases, different evolutionary
mechanisms intertwine and lead to scale-dependent feedback, manifested
by coevolution from genetic variations, spatio-temporal population
dynamics and spatially-varying selection pressure from the environment.

We will focus on a phage-microbe ecosystem, which is usually modeled
simply through Lotka-Volterra dynamics.  However, in the microbial
world, ecological relationships are more complicated than this due to
rapid evolution at the genomic scale.  In fact, it seems that phage are
multifunctional: now only do they exert predation pressure that reduces
the bacteria population, but they also transfer genes that can help
increase the bacteria population.  The way in which this happens in
detail is a possible instance of multi-level selection: at the level of
the individual bacteria, phage attack is a strong selection pressure.
But at the level of the community, there is an emergent fitness benefit
which allows the population to grow and even expand its range.

A remarkable example showing the significance of HGT-involved
multiscale feedback as a driving force for evolutionary complexity and
stability is the most-abundant phototropic organism,
\textit{Prochlorococcus spp.} \cite{chisholm2015}.  This
marine cyanobacterium experiences predation from cyanophages that,
surprisingly, were found to carry photosynthesis genes.  Phylogenetic
study showed that these genes had been horizontally transferred first
from cyanobacteria to cyanophages and back and forth multiple times
\cite{sullivan2006}. Interactions with cyanophage are assumed to be
important for the evolutionary pattern and diversity of
\textit{Prochlorococcus}. Specifically, \textit{Prochlorococcus}
exhibits niche stratification of two dominant ecotypes: the high
light-adapted ecotype near the sea surface evolved 150 million years
ago from the ancestral low light-adapted ecotype at the lower sea
level.  Due to the depth-dependent absorbtion spectrum of light, the
different ecotypes utilize distinct light intensities and spectra.

What were the environmental and genetic drivers of the evolution of the
high light-adapted ecotype?  \textit{Prochlorococcus} has a highly
streamlined genome, and lives at low density in a nutrient-deficient
environment.  Thus, the required spatial adaptations were the result of
novel genes that presumably were distributed through viral-mediated
HGT. Consistent with this interpretation, \textit{Prochlorococcus} does
not possess standard defense mechanisms against phage attack, such as
CRISPR or prophages (for restriction-modification, the situation is not
clear) \cite{chisholm2015}. It seems that their principle means of
defense against phage is modification of cell surface molecules that
prevent phage attachment. These molecules are expressed from genes that
have been rapidly modified through mutations and horizontal gene
transfer with other bacteria phyla. These genes reside in genomic
islands and constitute the majority of the genetic diversity
\cite{chisholm2015}.

In recent work, we have performed a calculation from a minimal
stochastic model to show how HGT leads to collective coevolution of the
bacteria and their phages, leading to the emergence of stratified
ecotypes in the euphotic zone \cite{shih_cyano}.  Through HGT from
bacteria, phages acquire both beneficial and inferior genes that are
responsible respectively for efficient and inefficient photosynthesis
in a certain environment. Since phages have a relatively higher
mutation rate, they create a rapidly evolving reservoir of genes for
the host bacteria.  On the other hand, bacteria with highly streamlined
genomes create a slowly evolving, stable repository of beneficial genes
for phages by filtering out inferior genes under selection. By carrying
and transferring beneficial photosynthesis genes, there is evidence that phage improve their fitness, e.g. by optimizing their burst size, by supplementing the host cell's metabolism
\cite{lindell2005photosynthesis}. In reality most mutations are neutral
or deleterious; but HGT is blind to this. Thus in HGT with host
bacteria, on a fast time scale, phages evolve deleterious mutations,
but can be rescued by bacteria whose genome preserves genes on a longer
time scale. Eventually bacteria and phage form a collective state,
enabling the rapid adaptation and range expansion to the environment
nearer the ocean's surface.  This emergent mutualism occurs despite the
intrinsic antagonism between bacteria and phages. In short, HGT-driven
collective coevolution provides a natural unified explanation for the
features of \textit{Prochlorococcus} system, including highly
streamlined small genome but huge pan-genome, lack of defense
mechanisms against viral attack, niche stratification of ecotypes and
phage predator carrying photosynthesis genes. It is expected that this
type of mechanism can appear in other spatially-stratified systems,
where genes that benefit the evolution of both host and parasite could
be present.

\medskip
\noindent\textit{Current and future challenges}

A generic framework to study rapid eco-evolutionary dynamics with
multi-scale feedback requires not only population dynamics and genetic
evolutionary mechanisms but also the understanding of the origin of
genetic variations. One related long-standing puzzle is: Did the
selected phenotypes already pre-exist or were encoded in the phenotypic
variation in the ecosystem prior to the selection, or do they arise
through stress-induced mutagenesis --- de novo mutations induced by
strong selection pressure at a higher rate (and perhaps at different
loci) \cite{fitzgerald2019mutation}? How does stress-induced
mutagenesis feed back into niche construction? A crucial role is played
by the genotype-phenotype map, but how is it influenced by selection in
spatially and temporally varying environments?

\medskip
\noindent\textit{Concluding remarks}

We have primarily focused on the rapid evolution of microbial
ecosystems, which occurs through the interplay between gene flow,
spatial variation, and feedbacks between the organisms in the ecosystem
and the physical characteristics of the environment. These phenomena
are critical to understanding such critical issues as the emergence of
antibiotic resistance \cite{von2016dissemination} and the ongoing
dynamics of global climate change
\cite{hutchins2017microorganisms,monroe2018ecoevolutionary,abs2019microbial}. Another
extreme example of ecological-evolutionary feedback is the growing
realization that the cancer tumor microenvironment provides a strong
source of heterogeneity that underlies the rapid evolution of
chemotherapy resistance \cite{korolev2014turning} and the emergence of
collective sensing and decision-making \cite{deisboeck2009collective}.
Theoretical modeling of these important classes of problem requires
explicit handling of spatial structure and demographic fluctuations. To
understand how scale-dependent ecological-evolutionary feedback drives
the spatio-temporal evolution of ecosystem structure is a truly grand
challenge that requires a trans-disciplinary approach to be
successful.

\ack
This material is partially supported by the Na-
tional Aeronautics and Space Administration through the
NASA Astrobiology Institute under Cooperative Agreement
No. NNA13AA91A issued through the Science Mission
Directorate.\\

\bibliographystyle{ieeetr}
\bibliography{rapid_evolution_PhysicalBiology_ref}

\end{document}